\documentclass[twocolumn,amsmath,nofootinbib]{revtex4} 

\usepackage{url}
\usepackage{amssymb}
\usepackage{amsfonts}
\usepackage{amsbsy}
\usepackage{graphicx}
\usepackage{hyperref}
\usepackage{array} 
\usepackage{multirow}

\newcolumntype{x}[1]{%
>{\centering\hspace{0pt}}p{#1}}%
\providecommand{\openone}{\leavevmode\hbox{\small1\kern-3.8pt\normalsize1}}

\def\etal{{\frenchspacing\it et al.}}

\def\etc{{\frenchspacing\it etc.}}

\def\spose#1{\hbox to 0pt{#1\hss}}
\def\simlt{\mathrel{\spose{\lower 3pt\hbox{$\mathchar"218$}}
   \raise 2.0pt\hbox{$\mathchar"13C$}}}
\def\simgt{\mathrel{\spose{\lower 3pt\hbox{$\mathchar"218$}}
     \raise 2.0pt\hbox{$\mathchar"13E$}}}
 \def\simpropto{\mathrel{\spose{\lower 3pt\hbox{$\mathchar"218$}}
     \raise 2.0pt\hbox{$\propto$}}}

\def\beq#1{\begin{equation}\label{#1}}
\def\eeq{\end{equation}}
\def\beqa#1{\begin{eqnarray}\label{#1}}
\def\eeqa{\end{eqnarray}}

\def\ed{\end{document}}

\def\rn{}
\def\nn#1 #2{#2. #1}				
\def\nnn#1 #2 #3{#2. #3. #1}			
\def\nnnn#1 #2 #3 #4{#2. #3. #4 #1}		
\def\nnnnn#1 #2 #3 #4 #5{#2. #3. #4 #5. #1}	
\def\dualand{ and\hbox{ }}				
\def\multiand{, and\hbox{ }}				
\def\rf#1;#2;#3;#4;#5 {{\frenchspacing\par\rn#1, #3 {\bf #4}, #5 (#2). \par}}
\def\rg#1;#2;#3;#4;#5;#6 {{\frenchspacing\par\rn#1, #3 {\bf #4}, #5 (#2). \par}}
\def\rfbook#1;#2;#3;#4;#5 {{\frenchspacing\par\rn#1, {\it #3} (#5, #4, #2).\par}}
\def\rfprep#1;#2;#3 {{\par\frenchspacing\rn#1, #3 (#2).\par}}
\def\rfproc#1;#2;#3;#4;#5;#6 {{\frenchspacing\par\rn#1 #2, in {\it #3}, ed. #4 (#5: #6)\par}}
\def\rfprocp#1;#2;#3;#4;#5;#6;#7 {{\frenchspacing\par\rn#1 #2, in {\it #3}, ed. #4 (#5: #6), p#7\par}}

\begin{document}
\pdfoptionalwaysusepdfpagebox=5


\title{Nuclear War from a Cosmic Perspective\footnote{Based on my talk at the symposium {\it The Dynamics of Possible Nuclear Extinction} held February 28--March 1 2015 at The New York Academy of Medicine: \url{http://totalwebcasting.com/view/?id=hcf}}}
\author{Max Tegmark}

\address{Dept.~of Physics \& MIT Kavli Institute, Massachusetts Institute of Technology, Cambridge, MA 02139}

\date{\today}

\vspace{10mm}

\begin{abstract}
I discuss the impact of computer progress on nuclear war policy, both by enabling more accurate nuclear winter simulations and by affecting the probability of war starting accidentally. I argue that from a cosmic perspective, humanity's track record of risk mitigation is inexcusably pathetic, jeopardizing the potential for life to flourish for billions of years. 
\end{abstract}

\maketitle

\begin{figure}[pbt]
\centerline{\includegraphics[width=90mm]{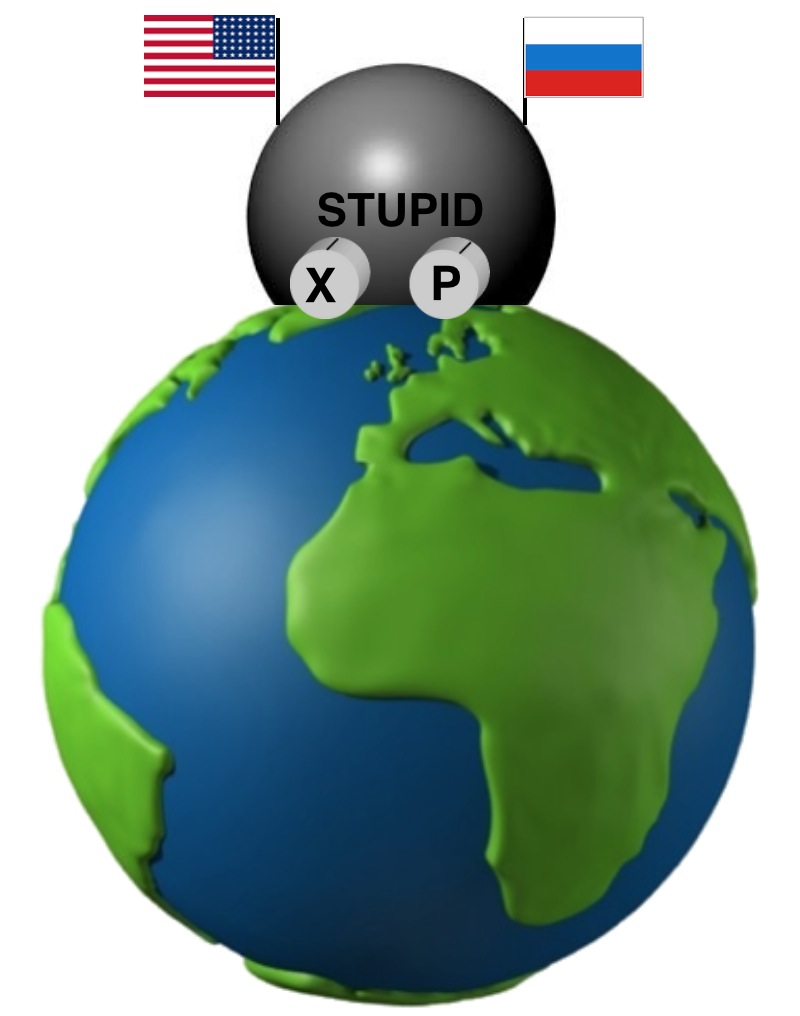}}
\caption{
We humans have invested great resources and ingenuity in building the \textit{
\textbf{S}pectacular
\textbf{T}hermonuclear
\textbf{U}npredictable 
\textbf{P}opulation
\textbf{I}ncineration
\textbf{D}evice}, 
 (acronym S.T.U.P.I.D.), whose two adjustable knobs determine its explosive power $X$ and the probability $P$ that it goes off spontaneously in any given year. }
\label{StupidFig}
\end{figure}

\section{S.T.U.P.I.D.}

13.8 billion years after our Big Bang, about 500 years after inventing the printing press, we humans decided to build a contraption called the \textit{
\textbf{S}pectacular
\textbf{T}hermonuclear
\textbf{U}npredictable 
\textbf{P}opulation
\textbf{I}ncineration
\textbf{D}evice}, 
abbreviated STUPID. 
It's arguably the the most costly device ever built on this beautiful spinning ball in space that we inhabit, but the cost hasn't prevented many people from saying that building and maintaining it was a good idea. This may seem odd, given that essentially nobody on our ball wants STUPID to ever get used.

It has only two knobs on the outside, labeled $X$ and $P$, but despite this apparent simplicity, it's actually a very complicated device. It's a bit like a Rube Goldberg machine inside, so complex that not a single person on our planet understands how 100\% of it works. Indeed, it was so complicated to build that it took the talents and resources of more than one country who worked really hard on it for many many years. Many of the world's top physicists and engineers worked to invent and build the technology for doing what this device does: creating massive explosions around the planet. But that was only part of the effort that went into it: to overcome human inhibitions towards triggering the explosions, STUPID also involves state-of-the-art social engineering, putting people in special uniforms and using peer pressure and the latest social coercion techniques to make people do things they normally wouldn't do. Fake alerts are created where people who refuse to follow missile launch protocols are fired and replaced, and so on.

\begin{figure*}[pbt]
\centerline{\includegraphics[width=204mm]{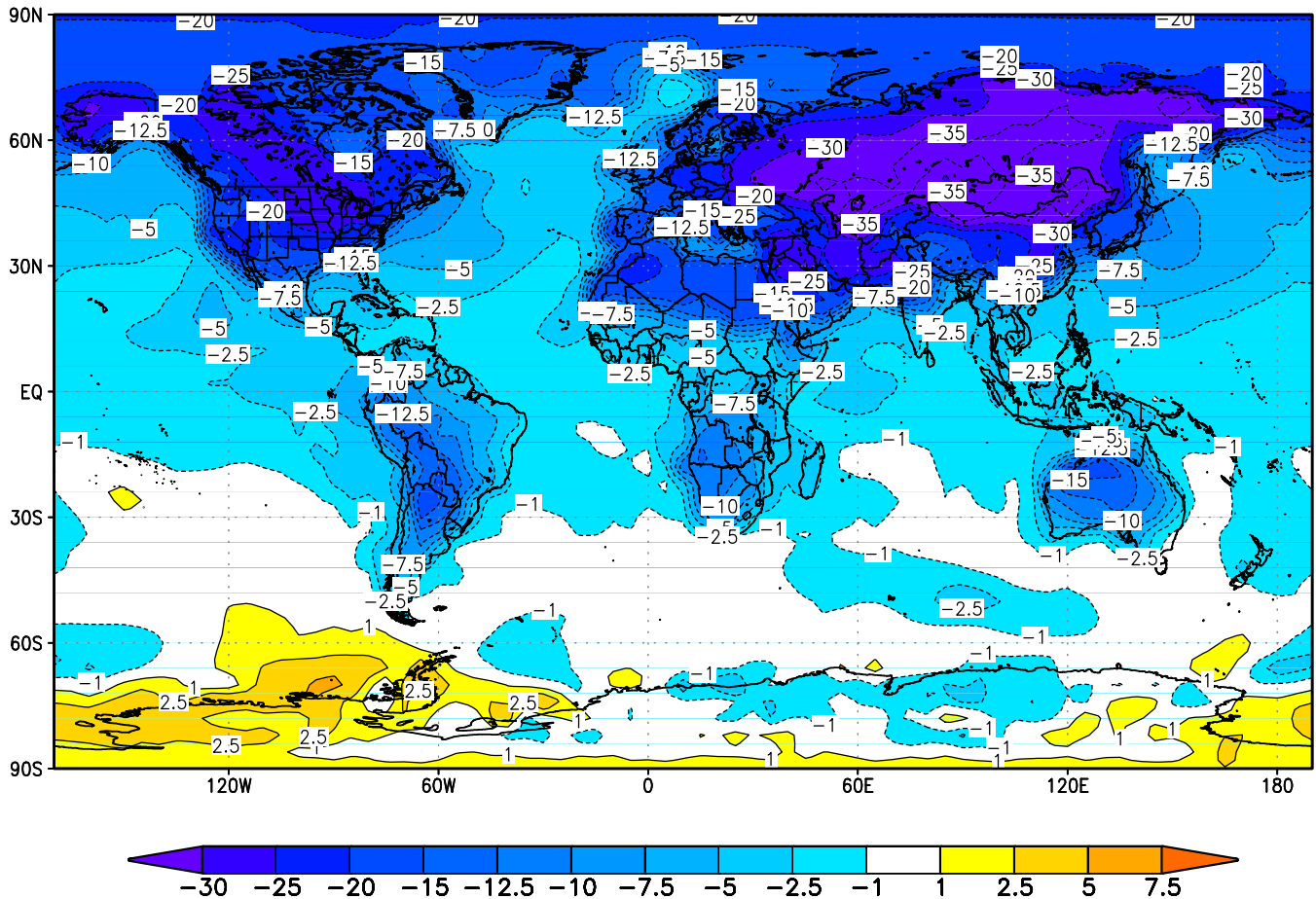}}
\vskip-10mm
\caption{
Average cooling (in $^\circ$C) during the first two summers after a full-scale nuclear war between the US and Russia. Reproduced with permission from \cite{Robock07}. 
}
\label{WinterFig}
\end{figure*}

Let's now focus on how STUPID works. What are these two knobs? The $X$-knob determines the total explosive power of the device. The $P$-knob determines the probability that this thing will go off during any random year for whatever reason. As we'll see, one of the nifty features of it is that it can spontaneously go off even if nobody wants it to. 

One can tune the settings of these two knobs, $X$ and $P$. Let's look a bit at how the setting of these two dials has evolved over time.  The $X$-knob was set to 0 until 1945, when we physicists figured out how to turn it up. We started below 20 kilotons with the Hiroshima and Nagasaki bombs, and by the time we got to the ``Tsar Bomba" in 1961, we were up to 50 megatons --- thousands of times more powerful.  The number of bombs also grew dramatically, peaking around 63,000 in the mid 1980's, dropping for a while and then holding steady around 16,000 hydrogen bombs in recent years, about 4,000 of which are on hair-trigger alert, meaning that they can be launched on a few minutes' notice \cite{Kristensen14}. 
Although those who decided to build STUPID argued that they had considered all factors and had everything under control, it eventually emerged that they had missed at least three crucial details.

\section{Nuclear Winter}

First, radiation risks had been underestimated, and over \$2Bn in compensation has been paid out to victims of radiation exposure from uranium handling and nuclear tests in the US alone \cite{Awards}. Second, it was discovered that using STUPID had the potential of causing a nuclear winter, which wasn't realized until about four decades after STUPID's inauguration---oops! Regardless of whose cities burned, massive amounts of smoke reaching the upper troposphere would spread around the globe, blocking out enough sunlight to transform summers into winters, much like when an asteroid or supervolcano caused a mass extinction in the past. When the alarm was sounded by both US and Soviet scientists in the 1980's \cite{Crutzen82,Turco83,Aleksandrov83,Robock84}, this contributed to the decision of Ronald Reagan and Mikhail Gorbachev to start turning town the $X$-knob. 

Today's climate models are significantly better than those run on the supercomputers of the 1980's, whose computational power was inferior to that of your smartphone. This enables more accurate nuclear winter forecasts.  Figure~2 (reproduced from \cite{Robock07}) shows the average change in surface air temperature (in degrees Celsius) during the first two summers after a full-scale nuclear war depositing 150 megatons of smoke into the upper troposphere. 

This calculation used a state-of-the-art general circulation model from NASA \cite{Schmidt06}, which includes a module to calculate the transport and removal of aerosol particles \cite{Koch06}, as well as a full ocean general circulation model with calculated sea ice, thus allowing the ocean to respond quickly at the surface and on yearly timescales in the deeper ocean. This was the first time that an atmosphere-ocean general circulation model was used in this context, and the first time that the time horizon was extended to a full decade. Unfortunately, the increased accuracy has revealed gloomier findings: Figure 1 shows cooling by about 20$^\circ$C (36$^\circ$ Fahrenheit) in much of the core farming regions of the US, Europe, Russia and China (by 35$^\circ$C in parts of Russia) for the first two summers, and about half that even a full decade later.

What does that mean in plain English? One doesn't need much farming experience to conclude that near-freezing summer temperatures for years would eliminate most of our food production. It's hard to predict exactly what would happen after thousands of Earth's largest cities are reduced to rubble and global infrastructure collapses, but whatever small fraction of all humans don't succumb to starvation, hypothermia or disease would need to cope with roving armed gangs desperate for food. 

Given the specter of Nuclear Winter, it has been argued that the traditional nuclear doctrine of Mutual Assured Destruction (MAD) has been replaced by Self-Assured Destruction (SAD) \cite{Robock12}: even if one of the two superpowers were able to launch its full nuclear arsenal against the other without any retaliation whatsoever, Nuclear Winter would assure its self-destruction.
Needless to say, there are many uncertainties in Nuclear Winter predictions, for example in how much smoke is produced and how high up it gets, which determines its longevity. Given this uncertainty, there is absolutely no basis for arguing that the $X$-knob is currently set low enough to guarantee the survival of most humans.

\section{Accidental Nuclear War}

Let's turn to the other knob, $P$: the probability that STUPID just goes kaboom for whatever reason. A third thing that the STUPID builders overlooked was that $P$ is set to an irrationally large value.  My own guess is that the most likely way we'll get a nuclear war going is by accident (which can also include people through various sorts of misunderstandings). We don't know what $P$ is and estimates vary widely. However, we know for sure that it's not zero, since there have been large numbers of close calls caused by all sorts of things: computer malfunction, power failure, faulty intelligence, navigation error, bomber crash, satellite explosion, {\etc} \cite{SchlosserBook}. In fact, if it weren't for heroic acts of certain individuals---for example Vasilii Arkhipov  and Stanislav Petrov---we might already have had a global nuclear war.  

What about the change of $P$ over time --- how has $P$ changed? Even though $P$ certainly dropped after 1990, when tensions subsided between the US and Russia, it might very well have gone up quite a bit again, and there are various reasons for this. The recent increase in mistrust and saber-rattling between the US and Russia obviously increases $P$, but there are also other seemingly unrelated developments that can potentially make $P$ larger. As just one small example among many that have been discussed, the US plan to replace 2 out of the 24 Trident submarine-launched ballistic missiles by conventional warheads, allegedly for potential use against North Korea, provides opportunities for misunderstanding.  An adversary seeing this missile coming and considering a nuclear response would have no way of knowing what kind of warhead it has.

Let me end talking about the impact of new technology on $P$, the risk of accidental nuclear war. Mutually Assured Destruction worked well when missiles were accurate enough to destroy a city but not accurate enough to destroy a silo. That made it very disadvantageous to launch any kind of first strike. Progress in computerized navigation has enabled much more precise targeting of missiles, reducing the disadvantage of a first strike, increasing $P$. Having accurate submarine-launched ballistic missiles near their targets also improves the prospects for a first strike. Most nuclear missile silos are within 2000 km of an ocean, from which submarine-launched ballistic missiles can destroy them in 7-13 minutes depending on how ``depressed" their trajectory is \cite{Gronlund92}. These shorter flight times give less time for the enemy to react, potentially making decision-makers jumpier, and as a result, both the US and Russia have now further increased $P$ by placing thousands of missiles on alleged hair-trigger alert, ready to launch on warning before a single nuclear explosion has been confirmed. 

What about artificial intelligence? There is broad consensus that artificial intelligence is now progressing rapidly. Although it is obviously very hard to forecast what will happen many decades from now, especially if AI turns out to surpass human cognitive abilities across the board, we can nonetheless draw some conclusions about likely developments in the near term as computers grow progressively more capable. For example, if we develop computer systems that are more reliable than people at properly following protocol, the military will have an almost irresistible temptation to implement them. We've already seen lots of the communications and command---and even analysis---be computerized in the military. Now, properly following proper protocol might sound like a pretty good thing, until you read about the Stanislav Petrov incident. Why was it, in 1983 when he got this alarm that the US was attacking the Soviet Union, that he decided not to pass it along to his superiors? Why did he decide not to follow proper protocol? Because he was human. If he had been a computer, he would have followed proper protocol, and some analysts speculate that a nuclear war might have started. 

Another concern is that the more we computerize decision making, the more we take what Kahneman calls ``system 1" out of the loop \cite{KahnemanBook}, and the more likely we are to lose valuable inhibitions and do dumb things. Suppose that president Putin had a person with him with whom he was friends, who carried the nuclear launch codes surgically implanted next to her heart. If the only way for him to get them was to first stab her to death, this might make him think twice before starting a nuclear war and jeopardizing billions of lives. If instead all he needs to do is press a button, there are fewer inhibitions. If you have a super-advanced artificial intelligence system that the president just delegates the decision to, the inhibitions are even weaker, because he's not actually authorizing launch: he's just delegating his authority to this system, deciding that if something happens in the future, then please go ahead and follow proper protocol. Given our poor human track-record of planning for the unforeseen (as illustrated in Kubrik's dark movie classic ``Dr. Strangelove"), I think that this would increase $P$.

Then there are good old bugs.  Has your computer ever given you the blue screen of death? Let's hope that the blue screen of death never turns into the red sky of death. Although it may be funny if it's just your unsaved work that got destroyed, it's less funny if it's your planet. 

Finally, another current trend seems to be that as AI systems get more and more advanced, they become more and more inscrutable black boxes where we just don't understand what reasoning they use --- but we still trust them.  The GPS in our car recently instructed me to drive down a remote forest road that ended in an enormous snow bank. I have no idea how it came to that conclusion, but I trusted it. If we have a super-advanced computer system which is telling the Russian military and political leadership that yes, there is an American missile attack happening right now, and here's the cool map with high resolution graphics showing the missiles, they might just trust it without knowing how it came to that conclusion.  If the system involved a human, they could ask it how it made that inference, and challenge its logic and input data, but if it was fully computerized, it might be harder to clear up misunderstandings before it was too late.

In summary, we don't know for sure that AI is going to increase the risk of accidental nuclear war, but we certainly can't say with confidence that it won't,  and it's very likely that the effects will be significant one way or the other. So it would be naive to think that the rise of artificial intelligence is going to have no impact on $P$.  

\section{Outlook}

Let me conclude by considering our place in a cosmic perspective. 13.8 billion years after our Big Bang, something remarkable has happened: life has evolved and our Universe has become aware of itself. This life has done many fantastic things that are truly inspiring. We've created great literature, music and film, and by using our curious minds we've been able to figure out more and more about our cosmos: How old it is, how grand it is and how beautiful it is. Through this understanding, we've also come to discover technologies that enable us to take more control and actually start shaping our destiny, giving us the opportunity to make life flourish far beyond what our ancestors had dreamt of. But we've also done some extremely dumb things here in our universe, such as building STUPID and leaving it running with its current knob settings. We don't yet know what effect AI and other future developments will have on the $P$-knob, but we can't rule out that things will get even worse. 

We professors are often forced to hand out grades, and if I were teaching Risk Management 101 and had to give us humans a midterm grade based on our existential risk management so far, you could argue that I should give a B- on the grounds that we're muddling along and still haven't dropped the course. From my cosmological perspective, however, I find our performance pathetic, and can't give more than a D: the long-term potential for life is literally astronomical, yet we humans are jeopardizing this future with STUPID, and devote such a tiny fraction of our attention to reducing $X$ and $P$ that this doesn't even become the leading election issue in any country. 

Why a D? Why not at least a B-, given that we're still not extinct? Many people view things from the traditional perspective that humans are the pinnacle of evolution, that life is limited to this planet, and that our focus should be limited to the next century or even just the next election cycle.  In this perspective, wiping ourselves out within a century may not seem like such a big deal. 

From a cosmic perspective however, that would be utterly moronic. It would be completely na\"{i}ve in a cosmic perspective to think that this is as good as it can possibly get. We have $10^{57}$ times more volume at our disposal. We don't have another century, but billions of years available for life to flourish. We have an incredible future opportunity that we stand to squander if we go extinct or in other ways screw up. People argue passionately about what the probability is that we wipe out in any given year: some guess it's 1\%, some guess much lower probabilities such as 0.0001, some guess higher.  Any of these numbers are just plain pathetic. If it's 1\% we'd expect to last of order a century, which is pretty far from the billions of years of potential that we have. Come on, let's be a little more ambitious here!

If you still have doubts about whether our priorities are faulty, ask yourself who is more famous: Vasili Arkhipov or Justin Bieber?  Then ask yourself which one of these two people should we thank for us all being alive today because his courageous actions may have singlehandedly stopped a Soviet nuclear attack during the Cuban Missile Crisis. 

The long-term survival of intelligent life on our planet is way too important to be left to leaders who have chosen to build and maintained STUPID.  Fortunately, history holds many examples of how a small number of idealistic individuals can make a large difference for the better. For example, according to both Reagan and Gorbachev, a major contributing factor to the deep nuclear cuts that they began in the 1980s was the research of that handful of scientists who discovered nuclear winter. There are many worthwhile efforts around the globe aimed at turning down $X$ and/or $P$. What can you personally do today to reduce the risk of nuclear apocalypse? Let me conclude by giving a concrete suggestion. I think that a strong and simple argument can be made that for any single country to have more than 200 nuclear weapons is unethical: 
\begin{enumerate}
\item Further increases in number cause negligible increases in deterrence: the deterrent effect on a potential attacker is already about as high as it can possibly get (please make a list of your 200 largest cities and imagine them suddenly obliterated), and when deployed on submarine-launched ballistic missiles, they are virtually immune to a surprise first-strike. 
\item  This is already at or above the threshold for causing a catastrophic global nuclear cold spell \cite{Robock07}, so increasing the number merely jeopardizes the future of humanity for no good reason. 
\end{enumerate}
If you accept this argument, then the logical conclusion is to stigmatize all efforts to replace or modernize nuclear weapons and any people or corporations that do so. The success in reducing smoking is an example to emulate. Why has the fraction of smokers in the US plummeted from 45\% in the 1950s to below 18\% today, most of whom say they would like to quit? Smoking hasn't been banned, but it has been stigmatized. In the 1950s, smoking was the cool thing to do, and movie stars and TV anchors all did it, whereas today's hip, rich and educated smoke much less than society's least fortunate members.  After scientists finally won the debate about whether smoking was harmful, the growing stigma caused ever more powerful organizations to work against it. Replacing or modernizing nuclear weapons is clearly worse for humanity than smoking, so ask yourself what you can do to dissuade companies from investing in it. For example, the non-profit organization ``Don't Bank on the Bomb" \cite{dontbank} provides all the information that you need to call your pension fund and encourage them to adopt a policy of not investing in nuclear weapons. If they ask you why, you can say ``I know that building nuclear weapons isn't illegal, but I don't want my money invested in it, just as I don't want it invested in tobacco, gambling or pornography". Many large banks, insurance companies and pension funds have already adopted such nuclear-free investment policies, and the momentum is growing. If quadruple-digit nuclear arsenals get the stigma they deserve and eventually become downsized, this of course won't eliminate the threat of nuclear war, but it will be a huge first step in the right direction.

I was invited to give this talk because I'm the president of the Future of Life Institute \cite{FLI}, a non-profit organization which we founded to help make humanity better stewards of this incredible opportunity we have to make life flourish for billions of years. All of us founders love technology --- every way in which 2015 is better than the stone age is because of technology. But we need to learn to handle technology wisely, and STUPID isn't wise---as Einstein put it: ``The splitting of the atom has changed everything except the way we think. Thus we drift towards unparalleled catastrophe." When we invented fire, we messed up repeatedly, then invented the fire extinguisher. With more powerful technologies such as nuclear weapons, synthetic biology and strong artificial intelligence, we should instead plan ahead and aim to get things right the first time, because it may be the only chance we'll get.Ê

I'm an optimist and believe that we often underestimate both what we can do in our personal lives and what life and intelligence can accomplish in our universe. This means that the brief history of intelligence so far is not the end of the story, but just the beginning of what I hope will be billions of years of life flourishing in the cosmos. Our future is a race between the growing power of our technology and the wisdom with which we use it: let's make sure that wisdom wins!

\bigskip

{\bf Acknowledgments:}
I wish to thank Helen Caldicott for inviting me to give the talk upon which this paper is based, 
Jesse Galef for  help transcribing it, Meia Chita-Tegmark for helpful feedback and Will Nelson for careful proofreading.

\end{document}